\documentstyle[aps]{revtex}
\begin{document}
\draft
\title{Static charged perfect fluid spheres in general relativity}
\author{B.V.Ivanov\thanks{%
E-mail: boyko@inrne.bas.bg}}
\address{Institute for Nuclear Research and Nuclear Energy,\\
Tzarigradsko Shausse 72, Sofia 1784, Bulgaria}
\maketitle

\begin{abstract}
Interior perfect fluid solutions for the Reissner-Nordstr\"om metric are
studied on the basis of a new classification scheme. It specifies which two
of the characteristics of the fluid are given functions and accordingly
picks up one of the three main field equations, the other two being
universal. General formulae are found for charged de Sitter solutions, the
case of constant energy component of the energy-momentum tensor, the case of
known pressure (including charged dust) and the case of linear equation of
state. Explicit new global solutions, mainly in elementary functions, are
given as illustrations. Known solutions are briefly reviewed and corrected.
\end{abstract}

\pacs{04.20.Jb}

\section{Introduction}

The unique exterior metric for a spherically symmetric charged distribution
of matter is the Reissner-Nordstr\"om solution. Interior regular charged
perfect fluid solutions are far from unique and have been studied by
different authors. The case of vanishing pressure (charged dust (CD)) has
received considerable attention. The general solution, in which the fluid
density equals the norm of the invariant charge density, was presented in
curvature coordinates by Bonnor \cite{one}. The proof that this relation
characterizes regular CD in equilibrium, i.e. in the general static case,
was given later \cite{two,three}. In the spherically symmetric case another
proof was proposed in Ref.\cite{four}. Concrete CD solutions were studied in
these coordinates \cite{four,five}. The generalization of the incompressible
Schwarzschild sphere to the charged case with constant $T_0^0$ was also
undertaken in a CD environment \cite{six}. Charged dust, however, has been
investigated more frequently in isotropic coordinates, since these encompass
the entire static case and allow to search for interior solutions to the
more general Majumdar-Papapetrou electrovacuum fields \cite{seven,eight}. In
both coordinate systems there is a simple functional relation between $%
g_{00} $ and the electrostatic potential. In isotropic coordinates there is
one non-linear main equation \cite{seven,nine} which has been given several
spherical \cite{ten,eleven,twelve,thirteen} and spheroidal \cite
{eleven,twelve,fourteen} solutions. One of them coincides with the general
static conformally flat CD solution \cite{fifteen}. These CD clouds may be
realized in practice by a slight ionization of neutral hydrogen, although
the necessary equilibrium is rather delicate. They have a number of
interesting properties: their mass and radius may be arbitrary, very large
redshifts are attainable, their exteriors can be made arbitrarily near to
the exterior of an extreme charged black hole. In the spheroidal case the
average density can be arbitrarily large, while for any given mass the
surface area can be arbitrarily small. When the junction radius $r_0$
shrinks to zero, many of their characteristics remain finite and
non-trivial. One can even entertain the idea for a point-like classical
model of electron were it not for the unrealistic ratio of charge $e$ to
mass $m$ \cite{one}.

Recently, new static CD solutions were found, in particular with density
which is constant or is concentrated on thin shells \cite{sixteen,seventeen}%
. In the spherically symmetric case a relation has been established with
solutions of the Sine-Gordon and the $\lambda \phi ^4$ equations \cite
{eighteen}.

The necessary condition for a quadratic Weyl-type relation has been derived
also for perfect fluids with non-vanishing pressure \cite{nineteen,twenty}.
However, in this case many other dependencies between the electrostatic and
the gravitational potential are possible, even when combined with constant $%
T_0^0$ \cite{twentyone}.

The original Schwarzschild idea of constant density has been also tested in
the charged case for a perfect fluid \cite
{twentytwo,twentythree,twentyfour,MB} or for imperfect fluid with two
different pressures \cite{one}. An electromagnetic mass model with vanishing
density has been proposed in Ref. \cite{six}. Unfortunately, the fluid has
negative pressure (tension). Although the junction conditions do not require
the vanishing of the density at the boundary, this is true for gaseous
spheres. A model with such type of density was proposed both in the
uncharged and the charged case \cite{twentyfive,twentysix}.

Another idea about the electromagnetic origin of the electron mass maintains
that, due to vacuum polarization, its interior has the equation of state $%
\rho +p=0$, where $\rho $ is the density and $p$ is the pressure. This leads
to tension, easier junction conditions and realistic $e$ and $m$ \cite
{twentyseven,twentyeight,twentynine,Ponce}. It can be combined with a
Weyl-type character of the field \cite{thirty}. The experimental evidence
that the electron's diameter is not larger than $10^{-16}$ cm, however,
requires that the classical models should contain regions of negative
density \cite{thirtyone,thirtytwo}. Probably an interior solution of the
Kerr-Newman metric is more adequate in this respect.

The presence of five unknown functions and just three essential field
equations allows one to specify the metric and solve for the fluid
characteristics \cite{thirtythree}. This is impossible in the uncharged
case. Another approach is to electrify some of the numerous uncharged
solutions. This has been done for one of the Kuchowicz solutions \cite
{thirtyfour} in Ref. \cite{thirtyfive}. Two other papers \cite
{thirtysix,thirtyseven} build upon the Wyman-Adler solution \cite
{thirtyeight,thirtynine}. Thus a charged solution is obtained, which has
approximately linear equation of state when $m/r_0$ is small. In Refs. \cite
{Pant,Tikekar,forty} generalizations of the Klein-Tolman (KT) solution \cite
{fortyone,fortytwo} were performed. Recently, static uncharged stars with
spatial geometry depending on a parameter \cite{fortythree,fortyfour} have
been generalized to the charged case \cite{fortyfive}. General
transformation connecting uncharged and charged solutions was introduced in
Ref. \cite{KN}.

The purpose of the present paper is to present a new and simple
classification scheme for charged static spherically symmetric perfect fluid
solutions. The calculations in each case are pushed as further as possible
and general formulae are given in many instances. The known solutions are
reviewed and compartmentalized according to this scheme in order to
illustrate general ideas, without being exhaustive. New solutions are added
where appropriate. The intention has been to stick to the simplest cases and
remain in the realm of elementary if not algebraic functions. The emphasis
is, however, on the general picture, which appears unexpectedly rich and
simpler than in the uncharged case.

The metric of a static spherically symmetric spacetime in curvature
coordinates reads 
\begin{equation}
ds^2=e^\nu dt^2-e^\lambda dr^2-r^2d\Omega ^2,  \label{one}
\end{equation}
where $d\Omega ^2$ is the metric on the two-sphere and $\nu ,\lambda $
depend on $r$. The fluid and its gravitation are described by five functions
depending on the radius: $\lambda ,\nu ,\rho ,p$ and the charge function $q$
which measures the charge within radius $r$. There are only three essential
field equations, hence, two of the above characteristics must be given. We
shall classify solutions according to this feature. For example, $\left( \nu
,\lambda \right) $ is the case of given metric and the other three fluid
characteristics are found from the equations. This does not mean that
solutions are distributed among groups which do not overlap. Thus, $\left(
\rho ,q\right) $ is a completely general case - any solution, after $\rho $
and $q$ are known, may be put into this class. The essence is that $\rho $
and $q^2$ are given functions and there is control over them; they can be
chosen regular, positive and comparatively simple. Then the other three
functions are usually more complex and are not always physically realistic.

In Sec.II the Einstein-Maxwell equations are organized into three main and
two auxiliary ones. Two of the main equations are universal for all cases,
the third one varies from case to case. Cases with given $\nu $ have linear
first-order differential equations for $e^{-\lambda }$. Cases with given $%
\lambda $ have linear second-order equations for $e^{\nu /2}$ or non-linear
first order equations for $\rho +p$. These results hold also for fluids with
a linear equation of state. The difficulties in the cases $\left( \rho
,p\right) $ and $\left( \rho ,q\right) $ are discussed, which prevent their
analytical treatment. The junction conditions are given in general form and
a reasonable set of physical requirements is included.

In the following table a summary is given of the discussed cases, the known
solutions which are reviewed and the new general and particular solutions.
They have been checked by computer for realistic properties.

\begin{table}[tbp]
\caption{Summary}
\begin{tabular}{llll}
Section & Cases & New solutions, Eqs. & Old solutions, Refs. \\ \hline
II & $\left( \rho ,p\right) ,\left( p,q\right) $ & none & none \\ 
III & $\left( \lambda ,\nu \right) ,\left( \lambda ,Y\right) $ & 36-8 & 21,35
\\ 
IV & $\left( \lambda ,n=-1\right) $ & 41-4, 49, 55 & 28,29,31,32 \\ 
V & $\left( \nu ,p\right) $ & 56-8, 75-8, 81, 84-7 & 1,4-6,10-13,15-18, 59
\\ 
VI & $\left( \nu ,q\right) $ & 92-4 & 38,39,63-67 \\ 
VII & $\left( \nu ,n\right) $ & 105-6, 108-10, 111-3, 114 & 36,53 \\ 
VIII & $\left( \nu ,\rho \right) $ & 117-8 & none \\ 
IX & $\left( \lambda ,\rho \right) ,\left( \lambda ,q\right) ,\left( \rho
,q\right) $ & 136-9, 145-6 & 6,22,23,25-27,37,49,64 \\ 
X & $\left( \lambda ,n\right) $ & 154, 156 & 42-44,77
\end{tabular}
\end{table}

The parameter $n$ designates the linear equation of state Eq. (17), which
represents a relation between $\rho $ and $p,$ while $Y\equiv \rho +p$.
Cases which contain $q$ are direct generalizations of uncharged solutions
depending usually on a parameter. After some reformulation the same property
is shared by the cases of charged dust $\left( \nu ,p=0\right) $ and
electrified de Sitter solutions $\left( \lambda ,n=-1\right) $.

Sec. XI contains some discussion and conclusions.

\section{Main equations and classification}

The Einstein-Maxwell equations are written as 
\begin{equation}
\kappa T_0^0\equiv \kappa \rho +\frac{q^2}{r^4}=\frac{\lambda ^{\prime }}r%
e^{-\lambda }+\frac 1{r^2}\left( 1-e^{-\lambda }\right) ,  \label{two}
\end{equation}
\begin{equation}
\kappa p-\frac{q^2}{r^4}=\frac{\nu ^{\prime }}re^{-\lambda }-\frac 1{r^2}%
\left( 1-e^{-\lambda }\right) ,  \label{three}
\end{equation}
\begin{equation}
\kappa p+\frac{q^2}{r^4}=e^{-\lambda }\left( \frac{\nu ^{\prime \prime }}2-%
\frac{\lambda ^{\prime }\nu ^{\prime }}4+\frac{\nu ^{\prime 2}}4+\frac{\nu
^{\prime }-\lambda ^{\prime }}{2r}\right) ,  \label{four}
\end{equation}
where the prime means a derivative with respect to $r$ and $\kappa =8\pi
G/c^4$. We shall use units where $G=c=1$. The charge function is obtained by
integrating the charge density $\sigma $. We shall use, however, $q$ as a
primary object and then 
\begin{equation}
\kappa \sigma =\frac{2q^{\prime }}{r^2}e^{-\lambda /2}.  \label{five}
\end{equation}
When $\sigma e^{\lambda /2}=const$, $q\sim r^3$, the case of the so-called
constant charge density. Spherical symmetry allows only a radial electric
field with potential $\phi $ given by 
\begin{equation}
F_{01}=\phi ^{\prime }=-\frac q{r^2}e^{\frac{\nu +\lambda }2}.  \label{six}
\end{equation}

Eq.(2) may be integrated by the introduction of the mass function 
\begin{equation}
M\left( r\right) =\frac \kappa 2\int_0^rr^2T_0^0dr=\frac \kappa 2%
\int_0^r\left( \rho +\frac{q^2}{\kappa r^4}\right) r^2dr,  \label{seven}
\end{equation}
and gives 
\begin{equation}
z\equiv e^{-\lambda }=1-\frac{2M}r.  \label{eight}
\end{equation}
This can be rewritten as 
\begin{equation}
\kappa \rho +\frac{q^2}{r^4}=\frac{2M^{\prime }}{r^2}=\frac 1{r^2}\left(
1-z-rz^{\prime }\right) ,  \label{nine}
\end{equation}
and constitutes the first of our main equations. The second is obtained as a
sum of Eqs. (2) and (3): 
\begin{equation}
\kappa \left( \rho +p\right) =\frac{e^{-\lambda }}r\left( \nu ^{\prime
}+\lambda ^{\prime }\right) =\frac zr\nu ^{\prime }-\frac{z^{\prime }}r.
\label{ten}
\end{equation}

The third main equation will not be Eq.(4) but another combination of
Eqs.(2)-(4) which varies from case to case. One can transform Eqs. (2)-(4)
into expressions for $p,q,\rho $%
\begin{equation}
2\kappa p=e^{-\lambda }\left( \frac{\nu ^{\prime \prime }}2-\frac{\lambda
^{\prime }\nu ^{\prime }}4+\frac{\nu ^{\prime 2}}4-\frac{\lambda ^{\prime }}{%
2r}+\frac{3\nu ^{\prime }}{2r}+\frac 1{r^2}\right) -\frac 1{r^2},
\label{eleven}
\end{equation}
\begin{equation}
\frac{2q^2}{r^4}=e^{-\lambda }\left( \frac{\nu ^{\prime \prime }}2-\frac{%
\lambda ^{\prime }\nu ^{\prime }}4+\frac{\nu ^{\prime 2}}4-\frac{\lambda
^{\prime }}{2r}-\frac{\nu ^{\prime }}{2r}-\frac 1{r^2}\right) +\frac 1{r^2},
\label{twelve}
\end{equation}
\begin{equation}
2\kappa \rho =e^{-\lambda }\left( -\frac{\nu ^{\prime \prime }}2+\frac{%
\lambda ^{\prime }\nu ^{\prime }}4-\frac{\nu ^{\prime 2}}4+\frac{5\lambda
^{\prime }}{4r}+\frac{\nu ^{\prime }}{2r}-\frac 1{r^2}\right) +\frac 1{r^2}.
\label{thirteen}
\end{equation}
These equations may be written as linear first-order equations for $z,$
suitable for the cases $\left( \nu ,q\right) ,\left( \nu ,p\right) $ and $%
\left( \nu ,\rho \right) $. Introducing $y=e^{\nu /2}$ we have 
\begin{equation}
\left( r^2y^{\prime }+ry\right) z^{\prime }=-2\left( r^2y^{\prime \prime
}-ry^{\prime }-y\right) z-2y+\frac{4q^2}{r^2}y,  \label{fourteen}
\end{equation}
\begin{equation}
\left( r^2y^{\prime }+5ry\right) z^{\prime }=-2\left( r^2y^{\prime \prime
}-ry^{\prime }+y\right) z+2y-4\kappa \rho r^2y,  \label{fifteen}
\end{equation}
\begin{equation}
\left( r^2y^{\prime }+ry\right) z^{\prime }=-2\left( r^2y^{\prime \prime
}+3ry^{\prime }+y\right) z+2y+4\kappa pr^2y.  \label{sixteen}
\end{equation}

In the uncharged case the prescription of an equation of state makes the
system of field equtions extremely difficult to solve. This is true even for
the simplest realistic linear equation of state 
\begin{equation}
p=n\rho -p_0,  \label{seventeen}
\end{equation}
where $n$ is a parameter taking values in the interval $[0,1]$ for
physically realistic solutions, while $p_0$ is a positive constant, allowing
the existence of a boundary of the fluid where $p=0$. When $p_0=0$ we obtain
the popular $\gamma $-law (with notation $n=\gamma -1$). In this case Eqs.
(2)-(4) with $q=0$ lead to an Abel differential equation of the second kind
independently from the approach or the coordinate system \cite{fortysix}. It
is soluble in few simple cases. Almost all of the few known solutions of the
more general Eq. (17) may be obtained by imposing a simple ansatz on $%
\lambda $ which makes the system overdetermined \cite{fortyseven}.
Therefore, it is surprising that in the more complex charged case fluids,
satisfying Eq. (17), are subjected to a linear equation, similar to Eqs.
(14)-(16). Plugging Eq. (17) into Eq. (10) and replacing the resulting
expression for $\rho $ in Eq. (15) yields 
\begin{equation}
\left( r^2y^{\prime }+\frac{5n+1}{n+1}ry\right) z^{\prime }=-2\left(
r^2y^{\prime \prime }+\frac{3-n}{n+1}ry^{\prime }+y\right) z+2y-\frac{%
4\kappa p_0}{n+1}r^2y.  \label{nineteen}
\end{equation}
We call this case $\left( \nu ,n\right) $. Eq. (18) was derived in Ref. \cite
{Humi} when $p_0=0$. Eq. (17) is a privileged one, due to its linearity.
Another realistic equation of state, namely the polytropic one, reads $%
p=n\rho ^{1+1/k}$. It causes the appearance of radicals in Eq. (15) and
leads to non-integrable equations.

Eqs. (14)-(16) and (18) may be written in the general form 
\begin{equation}
gz^{\prime }=f_1z+f_0,  \label{twentyone}
\end{equation}
whose quadrature is 
\begin{equation}
z=e^F\left( C+H\right) ,  \label{twentytwo}
\end{equation}
\begin{equation}
F=\int \frac{f_1}gdr\text{,\qquad }H=\int e^{-F}\frac{f_0}gdr.
\label{twentythree}
\end{equation}
Here and in the following $C$ will denote a generic integration constant.
They may be written also as linear second-order differential equations for $%
y $, useful in the cases $\left( \lambda ,q\right) ,\left( \lambda ,\rho
\right) ,\left( \lambda ,p\right) $ and $\left( \lambda ,n\right) $: 
\begin{equation}
2r^2zy^{\prime \prime }+\left( r^2z^{\prime }-2rz\right) y^{\prime }+\left(
rz^{\prime }-2z+2-\frac{4q^2}{r^2}\right) y=0,  \label{twentyfour}
\end{equation}
\begin{equation}
2r^2zy^{\prime \prime }+\left( r^2z^{\prime }-2rz\right) y^{\prime }+\left(
5rz^{\prime }+2z-2+4\kappa \rho r^2\right) y=0,  \label{twentyfive}
\end{equation}
\begin{equation}
2r^2zy^{\prime \prime }+\left( r^2z^{\prime }+6rz\right) y^{\prime }+\left(
rz^{\prime }+2z-2-4\kappa pr^2\right) y=0,  \label{twentysix}
\end{equation}
\begin{equation}
2r^2zy^{\prime \prime }+\left( r^2z^{\prime }+2\frac{3-n}{n+1}rz\right)
y^{\prime }+\left( \frac{5n+1}{n+1}rz^{\prime }+2z-2+\frac{4\kappa p_0}{n+1}%
r^2\right) y=0.  \label{twentyseven}
\end{equation}
The coefficient before the second derivative is one and the same in all
cases. Eq. (22) is the generalization to the charged case \cite
{thirtyfive,thirtyseven} of the Wyman equation \cite{thirtyeight}.
Obviously, the case $n=-1$ is not covered by Eqs. (18), (25).

One can find first-order differential equations also for the cases $\left(
\lambda ,*\right) $, based on the well-known Tolman-Oppenheimer-Volkoff
(TOV) equation \cite{fortytwo,fortyeight}, generalized to the charged case 
\cite{twentyeight}: 
\begin{equation}
\kappa p^{\prime }=-\kappa \left( \rho +p\right) \frac{2M+\kappa pr^3-q^2/r}{%
2r\left( r-2M\right) }+\frac{\left( q^2\right) ^{\prime }}{r^4}.
\label{twentynine}
\end{equation}
We can trade $q$ in Eq. (26) for $\rho $ and $\lambda $ by using Eq. (9).
The result is a Riccati equation for $Y$%
\begin{equation}
Y^{\prime }=-\frac r2e^\lambda Y^2+\frac{\lambda ^{\prime }}2Y-4\kappa \frac 
\rho r+\frac 2{r^4}\left( r^2M^{\prime }\right) ^{\prime }.  \label{thirty}
\end{equation}
It marks another way to solve the cases $\left( \lambda ,\rho \right)
,\left( \lambda ,n=-1\right) $. Its solution yields for the pressure $\kappa
p=Y-\kappa \rho $. Eq. (27) is a non-linear, but first-order companion of
Eq. (15). In somewhat different notation it was derived in Ref. \cite
{Tikekar}. Using Eq. (9) and the definition of $Y$ we find Riccati equations
for the cases $\left( \lambda ,q\right) $, $\left( \lambda ,p\right) $ and $%
\left( \lambda ,n\right) $ with the same coefficients before $Y^{\prime }$
and $Y^2$. In the last case it reads 
\begin{equation}
Y^{\prime }=-\frac r2e^\lambda Y^2+\left( \frac{\lambda ^{\prime }}2-\frac 4{%
\left( n+1\right) r}\right) Y-\frac{4\kappa p_0}{\left( n+1\right) r}+\frac 2%
{r^4}\left( r^2M^{\prime }\right) ^{\prime }.  \label{thirtythree}
\end{equation}
These equations may be transformed into linear second-order equations by a
change of variables from $Y$ to $y$ and we obtain exactly Eqs. (22)-(25). In
this process we have exchanged $Y$, which is the sum of the pressure and the
density, for $y$ which is part of the metric.

So far we have reformulated the original system of Eqs. (2)-(4) into Eqs.
(9), (10) and a third equation, presented in many different forms, adapted
to the cases of the proposed classification. We have briefly discussed the
cases $\left( \lambda ,\nu \right) ,\left( \lambda ,*\right) ,\left( \nu
,*\right) $. The three remaining cases $\left( \rho ,p\right) ,\left( \rho
,q\right) $ and $\left( p,q\right) $ are the most natural ones since one
prescribes two of the fluid characteristics, hoping that the third one and
the metric will be regular and reasonable. The case $\left( \rho ,q\right) $
is easily reduced to $\left( \lambda ,q\right) $ because of Eq. (9). In the
case $\left( \rho ,p\right) $, $Y$ is also known and Eq. (27) becomes an
intricate, non-linear, second-order equation for $M$ which is not simpler
than the TOV equation. It seems that it can be dealt with only numerically.

The third main equation for the case $\left( p,q\right) $ is obtained in the
following way. Let us replace the density in the TOV equation (26) with its
expression from Eq. (9). After some tedious manipulations, the following
equation for $M$ is found 
\begin{equation}
\left( M+g_0\right) M^{\prime }=f_2M+f_3,  \label{thirtyseven}
\end{equation}
whose coefficients are functions of $p$ and $q$. There is a standard
procedure for the solution of such equations \cite{fortysix,fortynine}. It
consists of two changes of variables which bring them to the Abel equation
of the second kind 
\begin{equation}
\omega \omega _\zeta -\omega =f\left( \zeta \right) ,  \label{fortyone}
\end{equation}
Its integrable cases are few, depend on the shape of $f\left( \zeta \right) $
and are tabulated in Ref. \cite{fortynine}.

As a whole, the most attractive are the mixed cases $\left( \nu ,*\right) $
where one fluid characteristic and one metric component are specified.

The five functions which describe the fluid together with its gravitational
field should satisfy some physical requirements. Eqs.(7)-(8) show that at
the centre $M\left( 0\right) =0$ and $e^\lambda =1$. The density and
pressure should be positive and monotonically decreasing towards the
boundary. It is obvious that $q^2$ should be positive, too. The boundary $r_0
$ of the fluid sphere is determined by the relation $p\left( r_0\right) =0$
where a junction to the Reissner-Nordstr\"om (RN) metric 
\begin{equation}
e^\nu =e^{-\lambda }=1-\frac{2m}r+\frac{e^2}{r^2}  \label{fortyfour}
\end{equation}
should be performed. The metric and $\nu ^{\prime }$ must be continuous
there. This leads to the expressions 
\begin{equation}
\frac m{r_0}=1-e^\nu \left( 1+\frac{r_0\nu ^{\prime }}2\right) =\frac{%
M\left( r_0\right) }{r_0}+\frac{q^2\left( r_0\right) }{2r_0^2},
\label{fortyfive}
\end{equation}
\begin{equation}
\frac{e^2}{r_0^2}=1-e^\nu \left( 1+r_0\nu ^{\prime }\right) =\frac{q^2\left(
r_0\right) }{r_0^2}.  \label{fortysix}
\end{equation}
The condition $e=q\left( r_0\right) $ follows from the vanishing of the
pressure and vice versa.

Solutions, satisfying the above conditions (which is already rather
non-trivial) are called in the following physically realistic. Some other,
more stringent requirements, are discussed in Refs. \cite
{fifty,fiftyone,fiftytwo}. The most important of them is $0\leq \left( \frac{%
dp}{d\rho }\right) ^{1/2}\leq 1$, which means that the speed of sound is
positive and causal. One has control over this characteristic in CD and
models with linear equation of state. The case $\left( \nu ,q\right) $,
which is a generalization of uncharged solutions, shares their behavior for
small $q$. In other cases the expressions for density and pressure may be so
complicated that the fulfilment of this requirement is hard to estimate.

\section{The cases $\left( \lambda ,\nu \right) $ and constant $T_0^0$}

In the case $\left( \lambda ,\nu \right) $ Eqs. (11)-(13) should be used to
find $q,p$ and $\rho $. This is the simplest case but control over pressure
and density is completely lost and one must proceed by trial and error.
Krori and Barua \cite{thirtythree} have given the solution 
\begin{equation}
\lambda =a_1r^2,\qquad \nu =a_2r^2+a_3,  \label{fortyeight}
\end{equation}
where $a_i$ denote generic constants of a known solution. They are fixed by
the junction conditions. The solution is non-singular and the positivity
conditions are satisfied.

Eqs.(7)-(8) show that the generalized Schwarzschild condition $T_0^0=const$
determines $\lambda $%
\begin{equation}
e^{-\lambda }=1-ar^2,  \label{fifty}
\end{equation}
where $a$ is a positive constant. We have the freedom to choose one more
function. In Ref. \cite{six} the condition $p=0$ was further imposed. We
shall review this solution in Sec.V. In Ref. \cite{twentyone} some relations
between $\phi $ and $\nu $ were utilized. They can be arbitrary in the
perfect fluid case, so we do not base the classification on such criteria.
One should use instead Eqs. (22)-(25) or Eq. (27). The simplest expressions
are obtained in the case $\left( \lambda ,Y\right) $ when Eq. (27) is used
to determine $\rho $%
\begin{equation}
4\kappa \rho =-rY^{\prime }-\frac{r^2Y^2}{2\left( 1-ar^2\right) }+\frac{arY}{%
1-ar^2}+12a.  \label{fiftyone}
\end{equation}
Then $\kappa p=Y-\kappa \rho $ and 
\begin{equation}
q^2=r^4\left( 3a-\kappa \rho \right) =\frac{r^5}4\left[ Y^{\prime }+\frac{%
rY\left( Y-2a\right) }{2\left( 1-ar^2\right) }\right] .  \label{fiftytwo}
\end{equation}
The function $\nu $ is determined by a simple integration from Eq. (10) 
\begin{equation}
\nu ^{\prime }=\frac r{1-ar^2}\left( Y-2a\right) .  \label{fiftythree}
\end{equation}
This is an advantage over the second-order equations (22)-(25). Eqs.
(35)-(38) solve the problem in a minimal algebraic way. Several positivity
conditions follow for the master function $Y$ and for $\rho $: $3a>\kappa
\rho ,Y>2a,Y>\kappa \rho ,Y^{\prime }<0$.

In order to illustrate how the above scheme works one may take any of the
solutions elaborated upon in Ref. \cite{twentyone}, extract $Y$ and check
how the above equations and inequalities are satisfied. The simplest case
has 
\begin{equation}
Y=2a+a_1\left( 1-ar^2\right) ^{1/2}.  \label{fiftyfour}
\end{equation}

Of course, in any of the cases $\left( \lambda ,*\right) $ we can study the
subcase given by Eq. (35). This will be done in the following sections.

A similar problem arises when $\nu $ is given. Which function should be
prescribed in addition in order to have the simplest algorithm for
generating solutions? Again, $Y$ is the best choice. This is seen by taking
the difference of Eqs. (23),(24) or reformulating Eq. (10) 
\begin{equation}
yz^{\prime }-2y^{\prime }z+ryY=0.  \label{fiftyfive}
\end{equation}
This equation is much simpler than any of Eqs. (14)-(16).

\section{Charged de Sitter solutions and their generalization}

The uncharged case is a special integrable case with linear equation of
state (17) and $n=-1,$ $p_0=0,$ which is equivalent to the de Sitter
solution. The charged case is also completely solvable. The third main
equation is Eq. (27) which becomes 
\begin{equation}
\kappa \rho =\frac{\left( r^2M^{\prime }\right) ^{\prime }}{2r^3}=\frac 1{%
2r^2}\left( 2M^{\prime }+rM^{\prime \prime }\right) .  \label{fiftysix}
\end{equation}
Eq. (10) gives $\nu =-\lambda $ This is a feature also of the exterior RN
solution. Eq. (9) yields 
\begin{equation}
q^2=-\frac{r^5}2\left( \frac{M^{\prime }}{r^2}\right) ^{\prime }=\frac r2%
\left( 2M^{\prime }-rM^{\prime \prime }\right) .  \label{fiftyseven}
\end{equation}
Thus, when $M$ is given, all other unknowns follow from simple formulae. Eq.
(42) allows also one to take $q$ as a basis: 
\begin{equation}
\kappa \rho =2a_0-\frac{q^2}{r^4}-4\int_0^r\frac{q^2}{r^5}dr,
\label{fiftyeight}
\end{equation}
\begin{equation}
M=-2\int_0^r\bar r^2\int_0^{\bar r}\frac{q^2}{\breve r^5}d\breve rd\bar r+%
\frac{a_0}3r^3.  \label{fiftynine}
\end{equation}
Here $a_0$ is some positive constant and clearly $q=r^{2+\varepsilon
}q_0\left( r\right) $ where $\varepsilon >0$ and $q_0\left( 0\right) =const$%
. Eqs. (43)-(44) demonstrate the process of 'electrification' of de Sitter
space. The bigger the charge function, the lower the density until some
point $r_0$ is reached where $\rho \left( r_0\right) =0$ and consequently
the fluid sphere acquires a boundary. We already know that $M>0$. Eqs.
(41)-(42) show that $\rho $ and $q^2$ are both positive only when $M^{\prime
}>0$ and $2M^{\prime }\geq r\left| M^{\prime \prime }\right| $. The equality
holds at the boundary, where $M^{\prime \prime }\left( r_0\right) <0$. The
charge and the mass of the solution follow from the junction conditions
(32),(33) 
\begin{equation}
e^2=2r_0^2M^{\prime }\left( r_0\right) ,\qquad m=M\left( r_0\right)
+r_0M^{\prime }\left( r_0\right) .  \label{sixty}
\end{equation}

Obviously, there is an abundance of solutions since $M$ and its first two
derivatives have to satisfy few simple inequalities. Four solutions are
known in the literature. The $T_0^0=const$ condition leads to $q=0$ in the
interior and consequently to the de Sitter solution, which has constant
density. One can introduce, however, a surface charge $\sigma \sim \delta
\left( r-r_0\right) $ which gives $q=\theta \left( r-r_0\right) e$, $\rho
=\rho _0$ and 
\begin{equation}
z=1-\frac \kappa 3\rho _0r^2.  \label{sixtytwo}
\end{equation}
This is the solution of Cohen and Cohen \cite{twentyseven,twentynine}.
Another solution \cite{twentyeight} has 
\begin{equation}
M=\frac{\kappa ^2}{360}\sigma _0^2r^3\left( 5a_1^2-2r^2\right) ,
\label{sixtythree}
\end{equation}
Here $\sigma _0$ is the constant charge density, while $a_1$ is related to $%
a_0=\frac{\kappa ^2}{24}\sigma _0^2a_1$. This is one of the electromagnetic
models of the electron. When $\sigma _0\rightarrow 0$ the limit is flat
spacetime. We have shown, however, that $a_0$ is not obliged to vanish when $%
q=0$ so that, in general, the case $\left( \lambda ,n=-1\right) $ is an
electric generalization of de Sitter spacetime. The natural generalization
of flat spacetime is the CD solution, discussed in the next section. Another
solution with $M$ similar to the one in Eq. (47) and satisfying the relation 
$\kappa p=q^2/r^4$ was given in Ref. \cite{Ponce}. It has zero total charge
and a simple expression for $\nu $, coming from Eq. (3).

A fourth solution has been found by Gautreau \cite{thirty}. It has 
\begin{equation}
2M=c_1^2r-\frac{c_1^2a_2^2}r\sin ^2\frac r{a_2},  \label{sixtyseven}
\end{equation}
and $q=-r^2\phi ^{\prime }$. Many other solutions are possible and we give
one simple realistic example, a variation of the solution in Ref.\cite
{twentyeight}: 
\begin{equation}
M=a\left( r^3-r^4\right) ,\qquad z=1-2ar^2+2ar^3  \label{seventyone}
\end{equation}
\begin{equation}
\kappa \rho =2a\left( 3-5r\right) ,\qquad q^2=2ar^5  \label{seventytwo}
\end{equation}
The junction at $r_0=3/5$ gives $m=ar_0^3$, $e^2=2ar_0^5$.

The case $Y=0$ is easy because Eq. (27) collapses into the simple relation
(41). A generalization can be made by taking Eq.(41) as a basis. Then $q$ is
given again by Eq. (42), while Eq. (27) becomes 
\begin{equation}
Y^{\prime }=-\frac r2e^\lambda Y^2+\frac{\lambda ^{\prime }}2Y.
\label{seventyfive}
\end{equation}
This is a Bernoulli equation and, unlike the Riccati equation, it is readily
soluble in quadratures. Its general solution gives an expression for the
pressure 
\begin{equation}
\kappa p=\frac{e^{\lambda /2}}{C+\frac 12\int e^{3\lambda /2}rdr}-\kappa
\rho ,  \label{seventysix}
\end{equation}
where $C^{-1}=Y\left( 0\right) $. When $C\rightarrow \infty $ we return to
the previous case $Y=0$. Fortunately, Eq. (10) can be integrated explicitly
too and a closed expression is obtained for $\nu $%
\begin{equation}
e^{\nu /2}=A^{-1}e^{-\lambda /2}\left( 1+\frac 1{2C}\int_0^re^{3\lambda
/2}rdr\right) ,  \label{seventyseven}
\end{equation}
\begin{equation}
A=1+\frac 1{2C}\int_0^{r_0}e^{3\lambda /2}rdr.  \label{seventyeight}
\end{equation}
The second equation follows from the junction conditions. Thus, every
function $M$ leads to two solutions: one with trivial $Y$ and one with
non-trivial $Y$, satisfying Eq. (51). The trivial solution has the
disadvantage of negative pressure. Eq.(52) suggests that in the non-trivial
case solutions with positive pressure may exist.

Let us take Eq. (47) with somewhat different constants 
\begin{equation}
M=br^3\left( 1-br^2\right) .  \label{seventynine}
\end{equation}
where $b>0$. When $x\equiv r^2<2/5b$ the density is positive and decreasing.
The pressure is positive at the centre when $1/4b<C<5/12b$. Let us choose $%
b=0.01$ and $C=40$. Then $p$ has a maximum at $x_0=0.3$ and a root at $%
x_0=2.5$. This is a semi-realistic interior solution with $e^2=5b^2x_0^3$
and mass given by Eq. (45).

\section{The case of given pressure. Charged dust}

In this section we discuss the cases $\left( \nu ,p\right) $ and briefly $%
\left( \lambda ,p\right) $. The pressure is considered a known positive
function which decreases monotonically outwards and vanishes at the boundary
of the fluid sphere. The simplest case is $p=0.$ This represents charged
dust. The case $\left( \nu ,p\right) $ is much easier since the third main
equation (16) is linear and first-order. Something more, in Eq. (19) $%
f_1=-2g^{\prime }$. This allows to obtain a compact expression for $z$%
\begin{equation}
z=\frac 1{\left( 1+\frac{r\nu ^{\prime }}2\right) ^2}+\frac C{r^2e^\nu
\left( 1+\frac{r\nu ^{\prime }}2\right) ^2}+\frac{4\kappa }{r^2e^\nu \left(
1+\frac{r\nu ^{\prime }}2\right) ^2}\int_0^r\left( 1+\frac{r\nu ^{\prime }}2%
\right) e^\nu r^3pdr.  \label{eightythree}
\end{equation}
The knowledge of $\nu $ allows to satisfy two of the junction conditions by
choosing a function continuous at $r_0$ together with its derivative.
Eqs.(9)-(10) provide an expression for $q$ in terms of the known $\nu ,p$
and with $z\left( \nu ,p\right) $ from Eq. (56) 
\begin{equation}
\frac{q^2}{r^2}=1-z\left( 1+r\nu ^{\prime }\right) +\kappa pr^2.
\label{eightyfour}
\end{equation}
The density follows from Eqs. (10) and (56) 
\begin{equation}
\kappa \rho =\frac{2z}{r^2}-\frac 2{r^2\left( 1+\frac{r\nu ^{\prime }}2%
\right) }+\frac{2z\nu ^{\prime }}r+\frac{z\nu ^{\prime }}{r\left( 1+\frac{%
r\nu ^{\prime }}2\right) }+\frac{z\nu ^{\prime \prime }}{1+\frac{r\nu
^{\prime }}2}-\frac{4\kappa p}{1+\frac{r\nu ^{\prime }}2}-\kappa p.
\label{eightysix}
\end{equation}
When $r\rightarrow 0$, $z\rightarrow 1$ and the third term on the right
produces a pole unless $\nu ^{\prime }\rightarrow 2\nu _0r$, which means $%
\nu ^{\prime \prime }\rightarrow 2\nu _0$. Then $r\nu ^{\prime }\rightarrow 0
$ and the poles in the first two terms cancel. The last two terms approach
negative constants. In order to compensate them and have a positive density
at the centre, $\nu _0$ must be a positive constant and the inequality $8\nu
_0>5\kappa p\left( 0\right) $ should hold. Another consequence is that $%
e^\nu $ is an increasing function in the vicinity of the centre. Since $%
e^{-\lambda }$ is a decreasing function, which starts from $1$ and equals $%
e^\nu $ at $r_0$, we have $e^\nu \left( 0\right) <1$. Eq. (56) shows that $z$
has a pole at $r=0$ unless $C=0$.

Now we can understand the physical meaning of the terms in $z$. The third
term represents the contribution from the pressure to the metric. When $p=0$%
, $z$ is still non-trivial and represents the general CD solution. It has a
pole at the beginning of the coordinates and should not be used there. When $%
C=0$ only the first term remains and this is the regular CD solution in
curvature coordinates \cite{one}. The intricate proofs that this is the most
general regular solution \cite{two,four} are replaced here by the obvious
condition $C=0$. The second term in Eq. (56) may be induced in principle by
the third if the lower limit of the integral is changed. The first term may
be absorbed by the third if one makes the shift $\kappa p\rightarrow \kappa
p+\frac 1{2r^2}$.

Let us discuss in detail first the case of regular CD. It provides an
excellent illustration of the classification scheme. In the case $\left( \nu
,p=0\right) $ Eqs. (5)-(6) and (56)-(58) give 
\begin{equation}
e^\lambda =\left( 1+\frac{r\nu ^{\prime }}2\right) ^2,\quad q=\frac{r^2\nu
^{\prime }}{2\left( 1+\frac{r\nu ^{\prime }}2\right) },  \label{eightyeight}
\end{equation}
\begin{equation}
\kappa \rho =\kappa \sigma =\frac{r\nu ^{\prime \prime }+2\nu ^{\prime }+%
\frac 12r\nu ^{\prime 2}}{r\left( 1+\frac{r\nu ^{\prime }}2\right) ^3},\quad
\phi =-e^{\nu /2}+\phi _0.  \label{ninety}
\end{equation}
For the case $\left( \lambda ,p=0\right) $ we should use Eq. (59) to express
everything via $\lambda $%
\begin{equation}
\nu ^{\prime }=\frac 2r\left( e^{\lambda /2}-1\right) ,\quad q=r\left(
1-e^{-\lambda /2}\right) ,  \label{ninetytwo}
\end{equation}
\begin{equation}
\kappa \rho =\kappa \sigma =\frac 2{r^2}\left( e^{-\lambda /2}-e^{-\lambda }+%
\frac r2\lambda ^{\prime }e^{-\lambda }\right) .  \label{ninetyfour}
\end{equation}
These formulae involve only the first derivative of $\lambda $. The case $%
\left( q,p=0\right) $ is explicitly solvable, unlike the general case $%
\left( q,p\right) $: 
\begin{equation}
e^\lambda =\frac{r^2}{\left( r-q\right) ^2},\quad M=q-\frac{q^2}{2r},
\label{ninetyfive}
\end{equation}
\begin{equation}
\nu ^{\prime }=\frac{2q}{r\left( r-q\right) },\quad \kappa \rho =\kappa
\sigma =\frac 2{r^3}\left( r-q\right) q^{\prime }.  \label{ninetyseven}
\end{equation}
The last formulae clearly demonstrate the electrification of flat spacetime,
which is the trivial dust solution in the uncharged case. If we put $p=0$ in
Eq. (29) we still obtain a complicated Abel equation. One can check that $M$%
, given by Eq. (63), satisfies it. The most complicated case is $\left( \rho
,p=0\right) $. Introducing $w$ via $e^\lambda =w^{-2}$,  Eq. (61) gives 
\begin{equation}
\nu ^{\prime }=\frac{2\left( 1-w\right) }{rw},\quad q=r\left( 1-w\right) ,
\label{h}
\end{equation}
while Eq. (62) becomes an equation for $w$%
\begin{equation}
rww^{\prime }=-w^2+w-\frac \kappa 2r^2\rho .  \label{htwo}
\end{equation}
This is an Abel differential equation like Eq. (29). The procedure for its
solution was described briefly after Eq. (29) and brings it to its canonical
form \cite{fortysix,fortynine}. The set of density profiles, leading to
integrable $w$ is very restricted.

Several explicit dust solutions in curvature coordinates have been given.
Efinger \cite{five} studied the simple case $e^{-\lambda }=4/9$ in the
presence of a cosmological constant. When it is zero we have $q=r/3$ and
singular $e^\nu =a_1r$, $\rho =\sigma =4/9\kappa r^2$, $F_{10}=\frac 12\sqrt{%
a_1}r^{-1/2}$. Florides illustrated his general discussion \cite{four} with
a power-law solution $q=a_2r^{a_3+3}$. Finally, in Ref.\cite{six} the case
of constant $T_0^0$ was discussed. It leads to Eq. (35) and its replacement
in Eqs. (61)-(62).

As was mentioned in the introduction, most work on CD has been done in
isotropic coordinates 
\begin{equation}
ds^2=U^{-2}dt^2-U^2\left( dr^2+r^2d\Omega ^2\right) ,  \label{hten}
\end{equation}
where the simple relation $\phi =\pm U^{-1}$ holds and the only equation to
be solved is 
\begin{equation}
U^{\prime \prime }+\frac 2rU^{\prime }=-\frac \kappa 2U^3\rho .
\label{heleven}
\end{equation}
When $U$ is given the density is readily determined from Eq. (68). The
function $U^{-2}=a_1r^2+a_2$ was used in Refs. \cite{ten,fifteen}. Another,
more general function 
\begin{equation}
U=1+\frac{a_3}{r_0}+\frac{a_3\left( r_0^k-r^k\right) }{kr_0^{k+1}}
\label{htwelve}
\end{equation}
was studied for $k=2$ in Ref. \cite{twelve}, for $k=4$ in Ref. \cite{eleven}
and for general $k$ in Ref. \cite{thirteen}. Recently, the function $U=a_4%
\frac{\sin a_5r}r$ was studied \cite{sixteen,seventeen}.

When $\rho $ is given, it is convenient to transform Eq. (68) into 
\begin{equation}
U_{XX}=-\frac{\kappa \rho }2\frac{U^3}{X^4},  \label{hthirteen}
\end{equation}
where $X=1/r$. We may consider $\rho \left( U\right) $ as a known function,
chosen to simplify this equation. One possibility, leading to Bessel
functions, is $\rho =a_6/U^2$ \cite{sixteen,seventeen}. Other choices lead
to two-dimensional integrable models, like the sine-Gordon equation \cite
{eighteen}. The metric (67) describes also the regular static CD with no
symmetry. Then the case of constant fluid density $\rho _0$ is soluble in $cn
$, one of the Jacobi elliptic functions, but the solution cannot be
spherically symmetric. In the spherically symmetric case Eq. (70) becomes an
Emden-Fowler equation, whose integrable cases are tabulated in Ref. \cite
{fortynine}. Eq. (70) is not among them.

Let us go back to curvature coordinates (1) which are more convenient when
one wants to study also the case of non-vanishing pressure. From the
considerations about the regularity of $\rho $, given after Eq. (58), it
follows that the simplest function $\nu $ would be $\nu =\nu _0r^2$. This
choice coincides with Eq. (34) and leads to exponential functions and
probably to the error function in Eq. (56). We shall choose a different
option 
\begin{equation}
e^\nu =\left( a+br^2\right) ^k,  \label{hfifteen}
\end{equation}
where $k$ is an integer, $0<a<1$ and $b>0$. This ansatz has been thoroughly
studied in the uncharged case \cite{fifty,fiftytwo}. In this section we take 
$k=1$. The CD solution build on it has 
\begin{equation}
z=\left( \frac{a+br^2}{a+2br^2}\right) ^2,\quad q=\frac{br^3}{a+2br^2},
\label{hsixteen}
\end{equation}
\begin{equation}
\kappa \rho =\frac{2b\left( a+br^2\right) \left( 3a+2br^2\right) }{\left(
a+2br^2\right) ^3}.  \label{heighteen}
\end{equation}
Density is monotonically decreasing. The junction conditions give 
\begin{equation}
e=\frac{br_0^3}{a+2br_0^2},\quad m=r_0\left( 1-a-2br_0^2\right) .
\label{hnineteen}
\end{equation}
The condition $z=e^\nu $ at $r_0$ supplies the relation $r_0^2=\left( 1-4a+%
\sqrt{1+8a}\right) /8b,$which ensures the obligatory $e=m$. The r.h.s. is
positive when $a<1$, which is the case.

The main disadvantage of regular CD solutions is that they possess a fixed
ratio $e/m$ which is unrealistic, especially for classical electron models,
and requires the extreme RN solution as an exterior. Eq. (56) tells that
when the point $r=0$ is excluded, general CD solutions are possible. They
have the following characteristics 
\begin{equation}
z=\left( 1+\frac{r\nu ^{\prime }}2\right) ^{-2}\left( 1+C\frac{e^{-\nu }}{r^2%
}\right) ,  \label{htwentytwo}
\end{equation}
\begin{equation}
\kappa \rho =\left( 1+\frac{r\nu ^{\prime }}2\right) ^{-2}\left( 1+C\frac{%
e^{-\nu }}{r^2}\right) \frac{\nu ^{\prime }}r-\frac{z^{\prime }}r,
\label{htwentythree}
\end{equation}
\begin{equation}
\frac{q^2}{r^2}=1-\frac{1+r\nu ^{\prime }}{\left( 1+\frac{r\nu ^{\prime }}2%
\right) ^2}\left( 1+C\frac{e^{-\nu }}{r^2}\right) ,  \label{htwentyfour}
\end{equation}
\begin{equation}
\phi ^{\prime }=-e^{\nu /2}\frac{\left[ r^4\nu ^{\prime 2}-4C\left( 1+r\nu
^{\prime }\right) e^{-\nu }\right] ^{1/2}}{2r\left( r^2+Ce^{-\nu }\right)
^{1/2}}.  \label{htwentyfive}
\end{equation}
When $C=0$ they reduce to Eqs. (59)-(60). What is the physical significance
of such solutions? Let us consider a core of perfect fluid, occupying a ball
with radius $r_0$, with given $\nu $ and $p$. Then $z$ is determined from
Eq. (56) with $C=0$: 
\begin{equation}
z=\frac 1{\left( 1+\frac{r\nu ^{\prime }}2\right) ^2}\left[ 1+\frac{C\left(
r\right) }{r^2e^\nu }\right] ,  \label{htwentysix}
\end{equation}
\begin{equation}
C\left( r\right) =4\kappa \int_0^r\left( 1+\frac{r\nu ^{\prime }}2\right)
e^\nu r^3pdr.  \label{htwentyseven}
\end{equation}
At the junction $z$ is equivalent to Eq. (75) where $C=C_0\equiv C\left(
r_0\right) $. An observer cannot understand whether the interior solution
consists of perfect fluid or general CD since the only imprint left by the
pressure distribution inside is a constant. In principle, the RN metric
should be taken as an exterior, but there are no obstacles to take a general
CD metric and postpone the junction to RN till another point $r_1>r_0$. Thus
we obtain a triple-layered model with a perfect fluid core up to $r_0$ (zone
I), a halo of general CD from $r_0$ to $r_1$ (zone II) and the RN solution
for $r>r_1$ (zone III). The metric component $z$ is given correspondingly by
Eq. (79), Eq. (75) with $C_0$ and Eq. (31). For this purpose we choose a
continuous, together with its derivative, function $\nu $ in the region $%
0<r<r_1$. At the first junction $z$ is also continuous, while the pressure
drops to zero. The density is finite. In zone II the pressure remains zero,
while the density continues to decrease. Finally, at $r_1$ the density also
drops to zero (perhaps with a jump) and a RN solution follows till infinity,
making the composite solution asymptotically flat. As seen from Eqs.
(76)-(77), the constant $C_0$ shifts $e/m$ from $1$, like a perfect fluid
solution, occupying zones I and II would do.

This picture can be backed by an explicit example with the ansatz (71).
Since the details should not depend on the form of $p\left( r\right) $ but
only on the value of $C$, we shall discuss first zone II and its junction
with zone III. In the interval $\left[ r_0,r_1\right] $ we have 
\begin{equation}
z=\frac{\left( a+br^2\right) \left[ C_0+r^2\left( a+br^2\right) \right] }{%
r^2\left( a+br^2\right) ^2},  \label{htwenteight}
\end{equation}
The total mass and charge are given by 
\begin{equation}
m=r_1\left( 1-a-2br_1^2\right) ,\quad e^2=r_1^2\left( 1-a-3br_1^2\right) .
\label{hthirtyone}
\end{equation}
The junction condition at $r_1$ gives 
\begin{equation}
C_0=m^2-e^2.  \label{hthirtythree}
\end{equation}
This relation clearly shows the effect of pressure in zone I on the
mass-charge ratio, not altered by the CD halo in zone II. The result in zone
I was found in Ref. \cite{six}.

An example of a physically realistic solution is given by $a=0.01,$ $%
r_1=4.013\sqrt{C_0}$, $r_0=3.588\sqrt{C_0}$, $b=0.075/\sqrt{C_0}$. This
proves the existence of CD solutions with $\left| e\right| /m\neq 1$. They
sustain the value of $C$, obtained in zone I from a perfect fluid solution
with positive pressure. In Ref. \cite{Eimerl} the same question was studied
in isotropic coordinates but the solutions have negative density in the
vicinity of the origin.

It seems that there are no solutions of the type $\left( \nu ,p>0\right) $
in the literature. It must be stressed that this case is completely general,
unlike the cases with constant $T_0^0$ or with $\rho +p=0$, discussed in the
previous sections. Every perfect fluid solution may be reformulated as a $%
\left( \nu ,p\right) $ case. Let us proceed with the ansatz (71). We base
the discussion on $C\left( r\right) $ as a fundamental object, hence, it is
convenient to select a more involved $p$ leading to a simple $C\left(
r\right) $ given by Eq. (80). Let us take 
\begin{equation}
\kappa p=\frac{p_0-p_1r^2}{a+2br^2}  \label{hthirtyseven}
\end{equation}
and then 
\begin{equation}
C\left( r\right) =r^4\left( p_0-\frac 23p_1r^2\right) ,  \label{hthirtyeight}
\end{equation}
\begin{equation}
z=\frac{\left( a+br^2\right) ^2}{\left( a+2br^2\right) ^2}\left[ 1+\frac{%
r^2\left( p_0-\frac 23p_1r^2\right) }{a+br^2}\right] ,  \label{hthirtynine}
\end{equation}
\begin{equation}
\frac{q^2}{r^2}=\frac{r^4}{\left( a+2br^2\right) ^2}\left( b-bp_0-\frac 13%
ap_1\right) ,  \label{hforty}
\end{equation}
where $z$ follows from Eq. (79) while $q$ is calculated from the general
expression (57). Here $p_0$ and $p_1$ are positive constants, connected by $%
p_0=p_1r_0^2$. The ansatz (71) yields $\nu _0=b/a$ and the condition $\rho
\left( 0\right) >0$ becomes $8b/5a>\kappa p\left( 0\right) =p_0/a$. This is
weaker than a necessary condition for positive $q^2$, $b>p_0$. At the
boundary the density is always positive. We have also $p_0=3C_0/r_0^4$ and
using the numerical values from the CD example, one can express $p_0$ and $%
p_1$ in terms of $C_0$, in addition to $b,r_0$ and $r_1$. It can be shown
that for $C>0.06$, $\rho $ and $q^2$ are positive and the solution is
physically realistic. Thus we have constructed explicitly a triple solution
of the type PF-CD-RN. This ends the discussion of case $\left( \nu ,p\right) 
$.

The case $\left( \lambda ,p\right) $ relies on Eq. (24). Even the simplest
constant $z$ leads to special functions or non-integrable equations for $y$
when $p$ is chosen physically realistic.

\section{The case $\left( \nu ,q\right) $}

The cases $\left( \nu ,q\right) $ and $\left( \lambda ,q\right) $ comprise
the most direct generalizations of the numerous uncharged solutions. Setting 
$q=0$ one obtains one of them with the chosen ansatz for $\nu $ or $\lambda $%
. Like in the previous section, the case $\left( \nu ,q\right) $ is much
simpler. The third main equation is Eq. (14). Now $f_1$ can be represented
as a linear combination of $g^{\prime },g$ and $y$. This is a general result
which holds for any of Eqs. (14)-(16), (18). In this particular case we have 
\begin{equation}
f_1=-2g^{\prime }+\frac{8g}r-4y,\quad h=\int \frac{\nu ^{\prime }dr}{1+\frac{%
r\nu ^{\prime }}2},  \label{hfortyone}
\end{equation}
\begin{equation}
e^F=\frac{r^2e^{2h}}{e^\nu \left( 1+\frac{r\nu ^{\prime }}2\right) ^2},
\label{hfortytwo}
\end{equation}
\begin{equation}
H=2\int e^\nu \left( 1+\frac{r\nu ^{\prime }}2\right) e^{-2h}r^{-3}\left( 
\frac{2q^2}{r^2}-1\right) dr,  \label{hfortythree}
\end{equation}
and $z$ is given by Eq. (20). We shall use the ansatz (71) where $k\geq 1$.
In the uncharged case $k=1$ leads to Tolman's solution IV \cite{fortytwo}.
The case $k=2$ was discussed first by Wyman \cite{thirtyeight} as an
extension of Tolman's solution VI and later was studied in detail by Adler 
\cite{thirtynine}. Solution with $k=3$ was given by Heintzmann \cite
{fiftythree} (see also Ref. \cite{fiftytwo}). The general class was studied
by Korkina \cite{fiftyfour}, but her only explicit solution was the one of
Heintzmann. Later Durgapal \cite{fiftyfive} studied in detail the cases $%
k=1-5$. All of them satisfy the physical criteria used in this paper, but
some have irregular behaviour of the speed of sound $\left( dp/d\rho \right)
^{1/2}$ and of $p/\rho $.

Going to the charged case we have 
\begin{equation}
e^2=q^2\left( x_0\right) ,\quad \frac m{r_0}=1+\frac{1+\left( k+1\right)
\tau x_0}{1+\left( 2k+1\right) \tau x_0}\left( \frac{e^2}{x_0}-1\right) ,
\label{hfortysix}
\end{equation}
where $\tau =b/a$, $x=r^2$. Let us consider first the case $k=1$. Then 
\begin{equation}
z=\frac{1+\tau x}{1+2\tau x}\left( 1+Cx+2xQ\right) ,\quad Q=\int_0^x\frac{q^2%
}{x^3}dx.  \label{hfortyeight}
\end{equation}
The pressure and the density are given by 
\begin{equation}
\kappa \rho =\frac{\tau -3C\left( 1+\tau x\right) }{1+2\tau x}+\frac{2\tau
\left( 1+Cx\right) }{\left( 1+2\tau x\right) ^2}-\frac{2\left( 1+\tau
x\right) }{1+2\tau x}\left( 3Q+\frac{2q^2}{x^2}\right) +\frac{4\tau xQ}{%
\left( 1+2\tau x\right) ^2},  \label{hfifty}
\end{equation}
\begin{equation}
\kappa p=\frac{\tau +C+3\tau Cx+2\left( 1+3\tau x\right) Q}{1+2\tau x}+\frac{%
q^2}{x^2}.  \label{hfiftyone}
\end{equation}
When $q=0$ these expressions coincide with the pressure and the density from
Refs. \cite{fortytwo,fiftyfive}. We have $\kappa \rho \left( 0\right)
=3\left( \tau -C\right) $. Let us choose $q^2=K^2x^3$. Then the condition $%
p\left( x_0\right) =0$ yields a negative expression for $C$ which means that 
$\rho \left( 0\right) $ is positive. The condition $z=e^\nu $ at $x_0$
defines $a$. Eq. (91) yields $e^2=K^2x_0^3$ and we can express $c,a,e,m$
through $x_0,\tau ,K$.

Let us consider next the case $k=2$. Then Eqs. (89)-(90) give 
\begin{equation}
z=1+Ce^F+e^F\int \frac{2\left( a+bx\right) q^2dx}{\left( a+3bx\right)
^{1/3}x^3},\quad e^F=\frac x{\left( a+3bx\right) ^{2/3}}.  \label{hfiftyfive}
\end{equation}
When $q=0$ this is the result of Adler who used the simplifying assumption $%
e^FH=1$ instead of the ansatz (71). When $q^2=K^2x^3$ we obtain 
\begin{equation}
z=1+\frac{Cx}{\left( a+3bx\right) ^{2/3}}+\frac{4a}{5b}K^2x+\frac 25K^2x^2.
\label{hfiftysix}
\end{equation}
This is the corrected result of Nduka \cite{thirtysix} who wrote the field
equations with an error in the sign of $q^2$. In all his results $K^2$
should be replaced by $-K^2$ and the conclusions correspondingly altered.

The ansatz (71) with $r^2$ replaced by $r$ or $r^3$ was also studied \cite
{Singh}. In another paper Singh and Yadav \cite{thirtyseven} simplified Eqs.
(14), (19) by demanding $f_1=-f_0$. Letting $q=Kr$ they obtain the Euler
equation. It has three types of solutions, all of which are singular at $r=0$%
. If $K=0$ the non-singular solution of Adler is again recovered.

Other solutions with singular $y$ were discussed in Refs. \cite
{Singh,Burch,Rago,Mak1}. In Ref. \cite{Rago} a charged version (with
constant charge density) of Tolman's solution V was obtained. Further
examples of $\left( \nu ,q\right) $ solutions are given in Refs. \cite
{Mak1,Mak2} but their properties are not examined.

\section{The case $\left( \nu ,n\right) $}

This is the case with given $\nu $ and the perfect fluid satisfying the
linear equation of state Eq. (17). Realistic solutions have $0<n\leq 1$
which means constant speed of sound and causal behavior. Surprisingly, the
charged case is much easier than the uncharged one. The third main equation
is Eq. (18) with $n\neq -1$. The functions $F$ and $H$ read 
\begin{equation}
e^F=\frac{r^{-\left( \alpha +1\right) }e^{\beta h_1}}{e^\nu \left( A+\frac{%
r\nu ^{\prime }}2\right) ^2},  \label{hfiftynine}
\end{equation}
\begin{equation}
H=2\int e^\nu \left( A+\frac{r\nu ^{\prime }}2\right) r^\alpha e^{-\beta
h_1}\left( 1-\frac{2\kappa p_0}{n+1}r^2\right) dr,  \label{hsixty}
\end{equation}
where 
\begin{equation}
h_1=\int \frac{\nu ^{\prime }dr}{A+\frac{r\nu ^{\prime }}2},\quad A=\frac{%
5n+1}{n+1},  \label{hsixtyone}
\end{equation}
\begin{equation}
\alpha =\frac{1-3n}{5n+1},\quad \beta =\frac{4n\left( 9n+1\right) }{\left(
n+1\right) \left( 5n+1\right) }.  \label{hsixtytwo}
\end{equation}
We shall study the range $0\leq n\leq \infty $ and some negative $n$. The
case $n=-1/5$ demands special treatment commented upon at the end of the
section. The case $n=-1$ was discussed in Sec. IV. Each of the coefficients $%
\alpha ,\beta $ and $A$ has the same limits when $n\rightarrow \pm \infty $.
When $0\leq n\leq \infty $ their ranges are $-3/5\leq \alpha \leq 1$, $1\leq
A\leq 5$, $0\leq \beta \leq 36/5$. Expressions (97), (98) are generic for
the cases $\left( \nu ,q\right) ,\left( \nu ,n\right) ,\left( \nu ,\rho
\right) $. The case $\left( \nu ,q\right) $, discussed in the previous
section, is obtained by putting $\alpha =-3$, $\beta =2$, $A=1$, changing
the sign of $H$ and an obvious replacement for $q$ instead of $p_0$. The
case $\left( \nu ,\rho \right) $ will be studied in the next section.

In the present case Eq. (100) shows that $\alpha +1$ is always positive for $%
n\in \left[ 0,\infty \right] $. Therefore, $z$ will have a pole unless $C=0$%
. Hence $z=e^FH.$ After $z$ is found, Eqs. (10),(17) allow to extract the
density from the metric 
\begin{equation}
\kappa \rho =\frac 1{n+1}\left[ \kappa p_0+\frac zr\left( \nu ^{\prime }-%
\frac{z^{\prime }}z\right) \right] ,  \label{hsixtysix}
\end{equation}
It is more convenient to express $p$ and $q$ from Eqs. (17), (57)
respectively and lay all the difficulty in the calculations on the density.
Plugging Eqs. (97), (98) into Eq. (101) we obtain 
\begin{eqnarray}
\left( n+1\right) \kappa \rho  &=&\kappa p_0+\frac{4\kappa p_0}{\left(
n+1\right) \left( A+\frac{r\nu ^{\prime }}2\right) }+\frac 2{r^2}\left[ 
\frac zA-\frac 1{A+\frac{r\nu ^{\prime }}2}\right] +\left( 2-\frac \beta {A+%
\frac{r\nu ^{\prime }}2}\right) \frac{z\nu ^{\prime }}r+  \nonumber \\
&&\ \ \ \frac{z\left( r\nu ^{\prime }\right) ^{\prime }}{r\left( A+\frac{%
r\nu ^{\prime }}2\right) },  \label{hsixtynine}
\end{eqnarray}
which is an analog of Eq. (58) and leads to similar conclusions for $\nu $.
The poles in the two terms in the square brackets cancel when $r\rightarrow 0
$ because $z\rightarrow 1$ and $\nu ^{\prime }\rightarrow 2\nu _0r$. This
condition makes $\rho $ a well-defined function. We shall try for $\nu $ an
ansatz which generalizes the one in Eq. (71) 
\begin{equation}
e^\nu =\left( a+br^s\right) ^k,  \label{hseventy}
\end{equation}
where $s\geq 2$ is not necessarily an integer. When $s>2$ it can be shown
that at the centre the relation $\rho \left( 0\right) +3p\left( 0\right) =0$
holds. This is exactly the equation of state for the Einstein static
universe (ESU) \cite{fifty,fiftytwo}. The pressure is negative at the
centre, which is unrealistic. Therefore we set $s=2$. The integral for $z$
is simplified considerably when 
\begin{equation}
n_k=\frac{k+5+4\sqrt{k\left( 2k+1\right) }}{31k-25}.  \label{hseventyfour}
\end{equation}
The first few values are $n_1\approx 2.15,$ $n_2\approx 0.53,$ $n_3\approx
0.39,$ $n_4=1/3$. The first is beyond the physical range, the fourth is an
exact number. One cannot probe the region $n<n_\infty \approx 0.21$ with
this method. A series of models is obtained, parameterized by $k$. They have 
\begin{equation}
z=\frac{\int \left( 1+t\right) ^{k-1}t^{\frac{\alpha +1}2-1}\left( 1-\frac 2{%
n+1}\mu t\right) dt}{t^{\frac{\alpha +1}2}\left( 1+t\right) ^{k-2}\left[
A+\left( A+k\right) t\right] },  \label{hseventynine}
\end{equation}
where $t=\tau r^2$, $\mu =\kappa p_0/\tau $. Eq. (102) becomes 
\begin{eqnarray}
\left( n+1\right) \frac{\kappa \rho }\tau  &=&\mu +\frac{4\mu \left(
1+t\right) }{\left( n+1\right) \left[ A+\left( A+k\right) t\right] }+\frac 2t%
\left[ \frac zA-\frac{1+t}{A+\left( A+k\right) t}\right] +  \label{heighty}
\\
&&\ \ \ \frac{2kz}{1+t}\left[ 2-\beta \frac{1+t}{A+\left( A+k\right) t}%
\right] +\frac{4kz}{\left( 1+t\right) \left[ A+\left( A+k\right) t\right] }.
\nonumber
\end{eqnarray}
Now additional terms contribute to $\rho \left( 0\right) $ and $p\left(
0\right) $ and they can be made positive by a proper choice of $\tau $ and $%
p_0$. The pressure is given by Eq. (17) and the charge function by Eq. (57)
which becomes 
\begin{equation}
\tau q^2=t\left[ 1-\frac{1+\left( 2k+1\right) t}{1+t}z+\frac{\kappa p}\tau
t\right] .  \label{heightyone}
\end{equation}

An interesting case is $k=4$ with $n_4=1/3$ exactly. When $p_0=0$ and $q=0$
this is the equation of state of pure incoherent radiation and the case is
not integrable \cite{fortysix}. This specific charged version, which is not
the only possible one, however, is integrable and assuming $\mu =2/3$ we
have 
\begin{equation}
z=\frac{1+\frac 23t-\frac 27t^3-\frac 19t^4}{\left( 1+t\right) ^2\left(
1+3t\right) },  \label{heightyfive}
\end{equation}
\begin{equation}
\frac{\kappa p}\tau =\frac{5+9t}{12\left( 1+3t\right) }+\frac{3\left(
3+7t\right) z}{2\left( 1+t\right) \left( 1+3t\right) }-\frac{\frac 73+3t+%
\frac 97t^2+\frac 19t^3}{4\left( 1+t\right) ^2\left( 1+3t\right) }-\frac 23,
\label{heightysix}
\end{equation}
\begin{equation}
\tau q^2=t\left( 1-\frac{1+9t}{1+t}z+\frac{\kappa p}\tau t\right) .
\label{heightyseven}
\end{equation}
The pressure is positive, monotonically decreases and vanishes at $t_0=0.54$%
. The charge density has a negative part. The junction conditions yield $%
a=0.3$, $b=0.16/r_0^2,$ $m=0.5r_0$, $e=0.45r_0$. We have $e/m<1$. Probably,
all models in the series have a range of $\mu $ where they are physically
realistic.

The method used to simplify Eq. (98) resembles the Korkina-Durgapal method,
which was extended to the charged case $\left( \nu ,q\right) $ in the
previous section. There are other ways to simplify the integral for $z$ even
when $k=1$. One of them is to take again $n=1/3$. Then $\alpha =0,$ $A=2$, $%
\beta =3/2,$ $\gamma =1/2$. We have 
\begin{equation}
z=\frac{1+t}{t^{1/2}\left( 2+3t\right) ^{3/2}}\left[ -\frac 14\left( \mu
-4+3\mu t\right) \sqrt{\left( 2+3t\right) t}+\frac{4+\mu }{12}\sqrt{3}\ln
\left( 3t+1+\sqrt{3t\left( 2+3t\right) }\right) \right] ,  \label{hninety}
\end{equation}
where the integration constant was chosen so that $z\left( 0\right) =1$. The
pressure and the charge are given by Eq. (17), (102) and (107) 
\begin{equation}
\frac{\kappa p}\tau =-\frac{3\left( 1+2t\right) }{4\left( 2+3t\right) }\mu +%
\frac{9z}{4\left( 2+3t\right) }+\frac 1{2t}\left( \frac z2-\frac{1+t}{2+3t}%
\right) ,  \label{hninetyone}
\end{equation}
\begin{equation}
\tau q^2=t\left( 1-\frac{1+3t}{1+t}z\right) +\frac{\kappa p}\tau t^2.
\label{hninetytwo}
\end{equation}
The pressure is well-behaved for any $\mu $ and vanishes at $t_0$. However, $%
q^2$ also has a zero, but at $t_1$, and becomes negative in a region where $%
t>t_1$. Unfortunately, for most values of $\mu $ the inequality $t_1<t_0$
holds, meaning that $q^2<0$ in some part of the interior. This is grossly
unrealistic. We have been unable to find by computer simulations any
realistic solution.

Another case with $k=1$ which leads to simple functions can be guessed from
Eq. (100). When $n=-1/9$, $\beta \equiv 0$. We have also $\alpha =3$, $A=1/2$%
, $\gamma =1$. Although the pressure is negative, it is worth being
discussed. We have 
\begin{equation}
z=\frac{1+t}{\left( 1+3t\right) ^2}\left[ 1+\left( 2-\frac 32\mu \right) t-%
\frac{27}8\mu t^2\right] .  \label{hninetythree}
\end{equation}
When $0<\mu <4/27$ the pressure monotonically increases, crosses the $t$
axis and forms a boundary of fluid with tension inside. Let us take $\mu =0$%
. The charge function $q^2$ is positive and increasing. The junction
conditions give $a=0.55,$ $b=0.11/r_0^2,$ $m=0.23r_0$, $e=0.35r_0$ where $%
t_0=0.2$. This solution resembles the classical models of the electron in
Sec.IV.

Finally, some words should be said about the case $n=-1/5$, which is really
very special. Let us choose the simple ansatz (71) for $\nu $. The case is
formally soluble but $z$ has poles at the centre and is ill-defined. When $%
p_0=0$ and $q=0$, this case is soluble \cite{fortysix}. It is connected by a
Buchdahl transformation to the de Sitter solution with $n=-1$ \cite
{fiftysix,fiftyseven} and is best expressed in isotropic coordinates. This
transformation, however, does not work for electrostatic fields and probably
is not applicable to the present case of a perfect fluid plus an
electrostatic field.

There are two papers which come near to the results of the present section.
A solution with singular $y,\rho $ and $p$ was found in Ref. \cite{Humi}
when $p_0=0$. Another one is contained in Ref. \cite{thirtysix} which
belongs to the $\left( \nu ,q\right) $ case and is mentioned in Sec.VI. When 
$m/r_0$ is small, the solution approximately satisfies Eq. (17) with $n=3/10$%
, a value which is rather close to $1/3$. Its connection to our solutions is
not clear.

A final remark is in order. It may be tempting to take the limit $%
e\rightarrow 0$ in one of the solutions obtained here and try to find
analytic expressions for the non-integrable at present cases of $\gamma $%
-law uncharged solutions \cite{fortysix}. However, for the ansatz (71) used
in this section, this leads to $r_0\rightarrow 0$ and to flat spacetime. In
the case of general $\nu $, it seems that the above limit will produce
charged solutions with zero total charge. This is suggested by the behavior
of $\sigma $ in some of the models, namely, its change of sign in the
interior. The proper non-charged limit seems to be $q\left( r\right) \equiv
0 $. Then, however, we obtain the field equations of the uncharged fluid.
Hence, nothing is gained in comparison with the three approaches described
in Ref.\cite{fortysix}. They lead to Abel differential equations of the
second kind with few integrable cases.

\section{The case $\left( \nu ,\rho \right) $}

This case arises as a special limit of $\left( \nu ,n\right) $ when $%
n\rightarrow \infty ,$ $p_0\rightarrow \infty $. One sees from Eq. (17) that
in this limit $\rho \rightarrow \rho _0=p_0/n$ which can be made a finite
constant. More generally, the third main equation (15) is obtained from Eq.
(18) plus the change $\frac{p_0}{n+1}\rightarrow \rho \left( r\right) $.
Eqs. (99)-(100) give $\alpha =-3/5,$ $A=5$, $\beta =36/5$. The constant $C$
in Eq. (20) is zero again in order to prevent a pole at the centre. Eqs.
(97)-(99) 
\begin{equation}
z=\frac{2e^{\frac{36}5h_3}}{r^{2/5}e^\nu \left( 5+\frac{r\nu ^{\prime }}2%
\right) ^2}\int e^\nu \left( 5+\frac{r\nu ^{\prime }}2\right) r^{-3/5}e^{-%
\frac{36}5h_3}\left( 1-2\kappa \rho r^2\right) dr,  \label{thfive}
\end{equation}
\begin{equation}
h_3=\int \frac{\nu ^{\prime }dr}{5+\frac{r\nu ^{\prime }}2}.  \label{thsix}
\end{equation}
The parameters $n$ and $p_0$ have been exchanged effectively for the density
function and in this respect Eq. (115) resembles Eqs. (89)-(90) from the $%
\left( \nu ,q\right) $ case. When $\rho $ is constant we obtain a
generalization of the Schwarzschild interior solution \cite{fiftyeight}. It
is much more complicated than the other generalization with constant $T_0^0$%
. As seen from Eq. (115), the case $\rho =0$ is non-trivial and leads to
purely electromagnetic mass models.

Let us try again the ansatz (103) for $\nu $. The method which lead to an
infinite series of realistic solutions in the $\left( \nu ,n\right) $ case
is rather useless here even when $s=2$. Therefore, we put $k=1.$ Then $%
\gamma =-1/5$ and 
\begin{equation}
z=\frac{1+t}{t^{1/5}\left( 5+6t\right) ^{4/5}}\int \frac{\left( 1-\frac{%
2\kappa \rho }\tau t\right) dt}{t^{4/5}\left( 5+6t\right) ^{1/5}}.
\label{thtwelve}
\end{equation}
When $\rho =\rho _0$ the integrals bring forth the hypergeometric function 
\begin{equation}
z=\frac{1+t}{\left( 1+\frac 65t\right) ^{4/5}}\left[ F\left( \frac 15,\frac 1%
5,\frac 65;-\frac{6t}5\right) -\frac{\kappa \rho _0}{3\tau }tF\left( \frac 15%
,\frac 65,\frac{11}5;-\frac{6t}5\right) \right] .  \label{ththirteen}
\end{equation}
The charge is given by Eq. (113), while the pressure reads 
\begin{equation}
\frac{\kappa p}\tau =\frac{2z}{1+t}-2z_t-\frac{\kappa \rho }\tau .
\label{thfourteen}
\end{equation}

Searching for an elementary solution, one may consider some sophisticated
density profile, leading to a simple integral in Eq. (117). Unfortunately we
have found only solutions with $q^2<0$.

\section{The cases $\left( \rho ,q\right) $, $\left( \lambda ,\rho \right) $
and $\left( \lambda ,q\right) $}

Having in mind the results in the previous section, it is no wonder that the
case $\left( \rho ,q\right) $, where there is control over the evasive
charge function, is much more popular in the literature. Eq. (9) easily
relates this case to $\left( \lambda ,\rho \right) $ and $\left( \lambda
,q\right) $, and Eq. (22) or (23) may be used as a third main equation. Like 
$\left( \nu ,q\right) $, these cases are direct generalizations of uncharged
cases. Eqs. (7),(8) define $z$ in terms of $\rho $ and $q$%
\begin{equation}
z=1-\frac \kappa r\int_0^r\rho r^2dr-\frac 1r\int_0^r\frac{q^2}{r^2}dr.
\label{thseventeen}
\end{equation}
Eq. (9) shows that $\rho $ is regular when $z-1\sim r^2$ for small $r$. We
accept the ansatz 
\begin{equation}
z=1-ar^2-br^4.  \label{theighteen}
\end{equation}
It follows from Eq. (120) when $q=Kr^l$ with $l=2,3$ and $\rho =\rho _0-\rho
_1r^2,$where $\rho _0,\rho _1$ are non-negative constants. The ansatz for $%
\rho $ goes back to Tolman's solution VII. When $l=2$ or $l=3$ we have
respectively 
\begin{equation}
z=1-\frac 13\left( \kappa \rho _0+K^2\right) x+\frac{\kappa \rho _1}5x^2,
\label{thtwenty}
\end{equation}
\begin{equation}
z=1-\frac{\kappa \rho _0}3x+\frac 15\left( \kappa \rho _1-K^2\right) x^2.
\label{thtwentyone}
\end{equation}
The total charge and mass are given by 
\begin{equation}
e=Kx_0^{3/2},  \label{thtwentytwo}
\end{equation}
\begin{equation}
m=\frac 12\left[ a+\left( b+K^2\right) x_0\right] x_0^{3/2}.
\label{thtwentythree}
\end{equation}
When $\rho _1\neq 0$ this is a model of a gaseous sphere, both density and
pressure vanish at the boundary \cite{twentyfive,twentysix}. We have $\rho
_0=\rho _1x_0$. When $\rho _1=0$, $\rho _0\neq 0$ this is a generalization
of the incompressible fluid sphere for $l=3$ \cite{twentytwo} or $l=2$ \cite
{twentythree}. Finally, when $\rho _1=\rho _0=0$ we come to a model with
electromagnetic mass \cite{six}. The solutions given by Eqs. (46)-(47) also
belong to this ansatz. The case $b\neq 0$ appears to be simpler, due to a
peculiarity in Eq. (22) and leads to elementary functions. Eq. (35)
demonstrates that the other case, $b=0$, belongs to the family of solutions
with constant $T_0^0$. Eq. (22) becomes in terms of $x$%
\begin{equation}
\left( 1-ax-bx^2\right) y_{xx}-\left( bx+\frac a2\right) y_x-\left( \frac b4+%
\frac{q^2}{2x^3}\right) y=0.  \label{thtwentyfour}
\end{equation}
The coefficient before $y$ is constant when $l=3$ (or $K=0$). Let us study
this case first. Eq. (123) gives 
\begin{equation}
d\equiv -\left( \frac b4+\frac{q^2}{2x^3}\right) =\frac 1{20}\left( \kappa
\rho _1-11K^2\right) .  \label{thtwentyfive}
\end{equation}
A change of variables brings Eq. (126) to 
\begin{equation}
y_{\xi \xi }+dy=0,  \label{thtwentysix}
\end{equation}
\begin{equation}
\xi =\int_0^x\frac{dx}{\sqrt{1-ax-bx^2}}.  \label{thtwentyseven}
\end{equation}
This integral has several different expressions, but many properties of the
solutions depend only on the fact that $\xi \left( 0\right) =0$, $\xi $ and $%
\xi _x$ are positive. Eq. (128) is easily solved and the last function to be
determined, the pressure, is given by Eq. (10). At the junction the pressure
vanishes when 
\begin{equation}
2y_\xi \left( x_0\right) =z_x\left( x_0\right)   \label{ththirty}
\end{equation}
is satisfied. The charge is given by Eq. (124) while the mass becomes 
\begin{equation}
m=\frac 12\left[ \frac{\kappa \rho _0}3+\left( 6K^2-\kappa \rho _1\right) 
\frac{x_0}5\right] x_0^{3/2}.  \label{ththirtyone}
\end{equation}
When $\rho _1=0$ the mass is positive. When $\rho _1\neq 0$ this assertion
still holds because then $\rho _1x_0=\rho _0$. The positivity of the mass
does not impose any conditions on the constants $\rho _0$, $x_0$ and $K$
which determine the solution. The pressure reads

\begin{equation}
\kappa p=4\sqrt{z}\frac{y_\xi }y-\frac 13\kappa \rho _0+\frac 15\left(
3\kappa \rho _1+2K^2\right) x.  \label{ththirtytwo}
\end{equation}
To be more concrete, let us introduce the three classes of solutions of Eq.
(128) for $d=0$, $d<0$ and $d>0$ respectively: 
\begin{equation}
y=C_1+C_2\xi ,  \label{ththirtyfour}
\end{equation}
\begin{equation}
y=C_1e^{-\sqrt{-d}\xi }+C_2e^{\sqrt{-d}\xi },  \label{ththirtyfive}
\end{equation}
\begin{equation}
y=C_1\sin \sqrt{d}\xi +C_2\cos \sqrt{d}\xi .  \label{ththirtysix}
\end{equation}
The conditions $y>0$ and $p\left( 0\right) >0$ lead to $C_1>0$, $C_2>0$ when 
$d\geq 0$ and to $C_2>0$, $C_2>C_1$ when $d<0$. Therefore, $e^\nu $
increases with $r$ until it meets $z$ which decreases. We have $y\left(
0\right) <1$.

1) Case $d=0$. Eq. (127) means $\kappa \rho _1=11K^2$, while Eq. (123) gives 
$b=-2K^2$. We also have $x_0=\frac{\kappa \rho _0}{11K^2}$and the solution
is determined by two constants, $\rho _0$ and $K$. Eq.(130) results in 
\begin{equation}
C_2=\frac 1{66}\kappa \rho _0,\quad C_1=-\frac 1{66}\kappa \rho _0\xi _0+%
\sqrt{z_0},  \label{ththirtyeight}
\end{equation}
\begin{equation}
z_0=1-\frac 53v^2\text{,\quad }v\equiv \frac{\kappa \rho _0}{11\left|
K\right| }  \label{thforty}
\end{equation}
and $y$ is given by Eq. (133). $C_1$ is real when $v\leq \sqrt{3/5}$. The
charge is given by Eq. (124) and the mass is $m=\frac 4{33}\kappa \rho
_0x_0^{3/2},$ leading to the ratio $\left| e\right| /m=3/4v\geq \sqrt{15}/4.$
Due to the negative $b$ Eq.(129) reads 
\begin{equation}
\xi =\frac 1{\sqrt{-b}}\ln \left| -\frac{a+2bx}{2\sqrt{-b}}+\sqrt{z}\right| -%
\frac 1{\sqrt{-b}}\ln \left| 1-\frac a{2\sqrt{-b}}\right| .
\label{thfortytwo}
\end{equation}
In Ref. \cite{twentysix} another expression was used, true only in a part of
the allowed interval for $v$. The pressure reads 
\begin{equation}
\kappa p=\frac{4C_2\sqrt{z}}{C_1+C_2\xi }-\frac 13\kappa \rho _0+7K^2x.
\label{thfortyfive}
\end{equation}
The condition $p\left( 0\right) >0$ is fulfilled when $C_1<2/11$. This holds
when $v$ varies near $\sqrt{3/5}=0.77.$ Thus, there are solutions with
positive pressure.

2) Case $d<0$. This means $\kappa \rho _1<11K^2$ and the limiting case $\rho
_1=0$ is possible. We shall discuss first the general case. The boundary
conditions fix the constants $C_{1,2}$. The pressure is given by Eq. (132) 
\begin{equation}
\kappa p=4\sqrt{-dz}\frac{C_2e^{\sqrt{-d}\xi }-C_1e^{-\sqrt{-d}\xi }}{C_2e^{%
\sqrt{-d}\xi }+C_1e^{-\sqrt{-d}\xi }}-\frac 13\kappa \rho _0+\frac 15\left(
3\kappa \rho _1+2K^2\right) x.  \label{thfortyeight}
\end{equation}
A necessary condition for positive pressure at the center is $C_2>C_1$,
satisfied when 
\begin{equation}
z_x\left( x_0\right) =-\frac{\kappa \rho _0}3-\frac 25\left( K^2-\kappa \rho
_1\right) x_0  \label{thfifty}
\end{equation}
is positive, which leads to $\kappa \rho _1>6K^2$. Simplifications are
possible when $d=b$ or $C_1=0$ but they don't give solutions with positive
pressure.

Let us discuss the subcase $\rho _1=0,$ i.e. $\rho =\rho _0$ \cite{MB}. Now
Eq. (141) shows that $z_x\left( x_0\right) <0$, therefore, $C_2<C_1$ and the
pressure is negative in this subcase, which was not realized in the above
reference. Eq. (124) holds as it is, while Eq. (125) reads 
\begin{equation}
m=\left( \frac 16\kappa \rho _0+\frac 35K^2x_0\right) x_0^{3/2}.
\label{thfiftyseven}
\end{equation}
We have $b=K^2/5>0$ and consequently $a^2+4b>0$. Then 
\begin{equation}
\xi =\frac 1{\sqrt{b}}\arcsin \frac{a+2bx}{\sqrt{a^2+4b}}-\frac 1{\sqrt{b}}%
\arcsin \frac a{\sqrt{a^2+4b}}.  \label{thfiftyeight}
\end{equation}
In Ref. \cite{twentysix} an incorrect expression was used. Formula (143)
allows to make connection with the results of Ref. \cite{six}, where in
addition $\rho _0=0$, so that the mass arises entirely from the
electrostatic field energy. Eq. (141) still shows that the pressure is
negative. Eqs. (124) and (142) give the ratio $\left| e\right| /m=5/3\left|
K\right| x_0.$ Now $a=0$ and $z$ is positive at the boundary when $\left|
e\right| /m>\sqrt{5}/3$. The second term in Eq. (143) vanishes and combining
it with Eq. (134) yields 
\begin{equation}
y=C_1\exp \left( -\sqrt{\frac{11}2}\arcsin \sqrt{b}x\right) +C_2\exp \left( 
\sqrt{\frac{11}2}\arcsin \sqrt{b}x\right) .  \label{thsixtytwo}
\end{equation}
The passage to $\arccos $ scales $C_{1,2}$ and interchanges their places,
giving Eq. (4.6) from Ref. \cite{six}.

3) Case $d>0$ . Now we have $\kappa \rho _1>11K^2$ which permits the limit $%
K=0$. The two cases do not differ essentially. The value of $z_x\left(
x_0\right) $ is positive, while $b<0$. Hence, $\xi $ is given by Eq. (138),
like in Ref. \cite{twentyfive}. The positivity condition is even more
complicated than in the previous case and simplifications do not seem
possible. This is true even for the uncharged case, discussed in the
mentioned reference. We shall remark only that the simple case $\kappa \rho
_1=\kappa \rho _0=x_0=1$, $K=0$ has positive pressure profile. For small
enough $K$, the charged case must behave the same way.

Let us study next the case $l=2$ when $z$ is given by Eq. (122). Eq. (126)
is not soluble unless $b\equiv -\kappa \rho _1/5=0$ meaning that the density
is constant. Then it becomes a particular case of the hypergeometric
equation 
\begin{equation}
\chi \left( \chi -1\right) y_{\chi \chi }+\left[ \left( \eta _1+\eta
_2+1\right) \chi -\eta _3\right] y_\chi +\eta _1\eta _2y=0,
\label{thsixtysix}
\end{equation}
where $\chi =ax$, $\eta _1+\eta _2+1=1/2$, $\eta _3=0$, $\eta _1\eta
_2=K^2/2a$. These relations give 
\begin{equation}
\eta _1=-\frac 14\pm \frac 14\sqrt{1-\frac{8K^2}a}  \label{thsixtyseven}
\end{equation}
and $\eta _{1,2}\in [-1/2,0)$. The reality of $\eta _1$ is ensured by $%
\kappa \rho _0>23K^2.$ Thus, in the charged case it is not possible to put $%
\rho =\rho _0=0$. This case was studied by Wilson \cite{twentythree} who did
not recognize the appearance of the hypergeometric function. He assumed that
the pressure was positive and developed a series expansion for $y$. A linear
combination between the two fundamental solutions of Eq. (145) is a
candidate for a regular metric. In this paper the emphasis is laid on
solutions in elementary functions, so we shall not pursue this issue
further, except in the case where $K=0$. Then Eq. (145) is easily solved 
\begin{equation}
e^\nu =\left[ 2\left( 1-ar^2\right) ^{1/2}+a_1\right] ^2.
\label{thseventythree}
\end{equation}
This is the expected Schwarzschild interior solution \cite{fiftyeight},
contrary to the claims in Ref. \cite{twentythree}.

A model for a superdense star with the ansatze 
\begin{equation}
z=\frac{1-a_2r^2}{1+a_1a_2r^2},\quad q=\frac{Ka_2r^3}{1+a_1a_2r^2},
\label{thseventfour}
\end{equation}
has been discussed in the uncharged case for $a_1=2$ \cite{sixty,sixtyone}, $%
a_1=7$ \cite{sixtytwo} and a set of discrete values for $a_1$ \cite
{fortythree}. Recently, it was studied for arbitrary $a_1$ both in the
uncharged \cite{fortyfour} and the charged case \cite{fortyfive}.
Introducing the variable 
\begin{equation}
\eta =\left( \frac{a_1}{a_1+1}\right) ^{1/2}\left( 1-a_2r^2\right) ^{1/2},
\label{thseventysix}
\end{equation}
one obtains 
\begin{equation}
\left( 1-\eta ^2\right) y_{\eta \eta \eta }-\eta y_{\eta \eta }+d_1y_\eta =0,
\label{thseventyseven}
\end{equation}
\begin{equation}
d_1=a_1+2-\frac{2K^2}{a_1},  \label{thseventyeight}
\end{equation}
where $d_1>0$ is required.

Formally, Eq. (150) is a subcase of Eq. (126) with $a=0$, $b=1$, $%
x\rightarrow r$, $y\rightarrow y_\eta $, so we can use the machinery
developed there. We have $\xi =\arcsin \eta $. We may change the variable to 
$\delta =\arccos \eta $ because Eq. (128) is invariant under this change.
Hence, $y_\eta $ is given by Eq. (135) and $y$ may be found by integration.
Use of trigonometric equalities helps to find an expression with two terms 
\begin{equation}
y=a_4\left\{ \frac{\cos \left[ \left( \sqrt{d_1}+1\right) \delta +a_3\right] 
}{\sqrt{d_1}+1}-\frac{\cos \left[ \left( \sqrt{d_1}-1\right) \delta
+a_3\right] }{\sqrt{d_1}-1}\right\} ,  \label{thseventynine}
\end{equation}
This formula was found in Refs. \cite{fortyfour,fortyfive} by resorting
first to Gegenbauer functions and then to Chebyshev polynomials. The
peculiarities of Eq. (126) permit a straightforward derivation in elementary
functions.

Let us discuss next a solution of Nduka \cite{thirtyfive}, belonging to the
class $\left( \lambda ,q\right) $. He chose a constant $z\neq 1$ and
generalized the uncharged solution of Ref. \cite{thirtyfour} by taking $l=1$%
, i.e. $q=Kr$. Then Eq. (22) turns into the Euler equation and $y$ and $\rho 
$ are singular at $r=0$.

The ansatz (121) was examined also in Ref. \cite{Burch} where besides $q$ a
generating function $G$ was given, satisfying the relation $G^2zq=const.$
Therefore this is again the case $\left( \lambda ,q\right) $. Several
solutions were found, including the one of Nduka \cite{thirtysix} and a
singular in $y$ solution, which is a generalization of the solutions in
Refs. \cite{Pant,Whitman}.

\section{The case $\left( \lambda ,n\right) $}

In this case the fluid satisfies the linear equation of state (17) and some
ansatz for $\lambda $ is given. The other metric component is found from Eq.
(25) - a linear second-order equation for $y$.

Let us study first the case $z=c<1$ which leads to singular pressure and
density, but provides the opportunity to generalize the well-known KT
solution \cite{fortyone,fortytwo,fortysix,sixtythree,sixtyfour}. Eq. (25)
becomes 
\begin{equation}
r^2y^{\prime \prime }+k_0ry^{\prime }+\left( -k_1+k_2r^2\right) y=0
\label{theightyone}
\end{equation}
\[
k_0=\frac{3-n}{n+1}\text{,\quad }k_1=\frac{1-c}c,\quad k_2=\frac{2\kappa p_0%
}{\left( n+1\right) c}. \label{theightyone} 
\]
Its solution is given by Bessel functions \cite{fortynine} 
\begin{equation}
y=r^{\frac{1-k_0}2}\left[ C_1J_{k_3}\left( \sqrt{k_2}r\right)
+C_2Y_{k_3}\left( \sqrt{k_2}r\right) \right] .  \label{theightynine}
\end{equation}
\[
k_3=\frac 12\left[ \left( 1-k_0\right) ^2+4k_1\right] ^{1/2}. 
\]
When $p_0=0$ this becomes 
\begin{equation}
y^2=C_1r^{1-k_0+2k_3},  \label{theightytwo}
\end{equation}
It can be checked that $1-k_0+2k_3>0$. We have included only the
non-divergent at $r=0$ solution. The density and the charge function are
singular at the origin. This solution was derived in Ref. \cite{Pant} and
rediscovered in Ref. \cite{Herrera}. It was generalized by Tikekar \cite
{Tikekar} whose solution belongs to the $\left( Y,n\right) $ case and is
singular at the origin too. Another generalization was offered in Ref. \cite
{forty} but it alters many of its properties, including the equation of
state.

Let us discuss finally the ansatz (35), imposed by the condition of constant 
$T_0^0$, like in Sec. III. Introducing $x=ar^2$ turns Eq. (25) into 
\begin{equation}
x\left( x-1\right) y_{xx}+\left[ \frac{n+5}{2\left( n+1\right) }x-\frac 2{n+1%
}\right] y_x+Dy=0,  \label{thninety}
\end{equation}
\begin{equation}
D=\frac 1{2\left( n+1\right) }\left( 3n+1-\frac{\kappa p_0}a\right) .
\label{thninetyone}
\end{equation}
This is once again the hypergeometric equation and its solutions depend on
the sign of $D$. Particular cases are the de Sitter solution and ESU. There
are many elementary solutions but they have either singularities or it is
impossible to ensure that both $p$ and $q^2$ are positive.

\section{Discussion and conclusions}

Charged static perfect fluids have attracted considerably less attention
than the uncharged ones, the number of papers being roughly an order of
magnitude smaller. The introduction hints how diverse were the approaches to
their study. Charged dust occupies the first place in popularity. The rest
is a mixture of generalizations: of the Schwarzschild idea about
incompressibility, with the limiting case of vanishing density and
electromagnetic mass models, generalization of the idea about vacuum
polarization which brings forth the de Sitter solution, generalization
simply of well-known uncharged solutions like those of Adler, Kuchowicz,
Klein, Mehra, Vaidya-Tikekar and others, generalization of Weyl type
connections, well studied in the electrovac case.

In this paper we have tried to show that the charged case has a life of its
own when subjected to a natural classification scheme. Surprisingly, in many
respects it looks simpler than the uncharged case. The presence of the
charge function serves as a safety valve, which absorbs much of the fine
tuning, necessary in the uncharged case. The general formulae derived here
show that the abundance of solutions is probably bigger than in the
traditional case. The proposed scheme becomes rather trivial there,
representing ansatze mainly for $\lambda $, $\nu $ and sometimes $\rho $. In
the charged case, however, it allows to sort out the different ideas
mentioned above. Thus constant $T_0^0$ leads to the cases $\left( \lambda
=1-ar^2,*\right) $, while constant density - to the much more difficult
cases $\left( \nu ,\rho \right) $ and $\left( \lambda ,\rho \right) $,
similar to $\left( \nu ,n\right) $ and $\left( \lambda ,n\right) $.
Electromagnetic mass models are subcases, often spoiled by negative
pressure. The point-like idea seems to be viable only for CD, leading to
flat spacetime when the pressure does not vanish. The generalization of
uncharged solutions firmly occupies the cases $\left( \lambda ,q\right) ,$ $%
\left( \rho ,q\right) $ and $\left( \nu ,q\right) $. Models with $n=-1$ are
intimately related to the soluble case $n\neq -1$ and are much richer than
their traditional protagonist - the de Sitter solution. The other models
with negative pressure, which seem worth being studied, are the
generalizations of ESU with $n=-1/3$. Finally, the Weyl type relations, so
successful in electrovac and CD environments, seem rather out of place in a
'pressurized' perfect fluid.

Another advantage of this classification is that it delineates the degree of
difficulty and the most convenient points for attack of the problem. The
easiest cases seem to be $\left( \lambda ,Y\right) $ and $\left( \nu
,Y\right) $. If we insist on authentic fluid characteristics and not on
their combinations, then $\left( \nu ,p\right) $ and $\left( \nu ,q\right) $
are the best candidates. On the other hand, the most difficult are $\left(
\rho ,p\right) $ and $\left( p,q\right) $, which seem to be accessible only
numerically. The most unpredictable case is probably $\left( \lambda ,\nu
\right) ,$ since there is no control on exactly those characteristics which
must satisfy the majority of regularity and positivity conditions.

A different division is between general and special cases. As was noticed in
the introduction, a known solution belongs to any of the general cases; a
solution of $\left( \nu ,q\right) $ may be written as $\left( \lambda
,p\right) $ or even as $\left( \rho ,p\right) $ solution. Only the
simplicity of two of the five functions $\nu ,\lambda ,\rho ,p,q$ betrays
where was the starting point. The special cases that we discussed, apart
from the ansatz (35), are essentially three: $\rho +p=0$ and $\left( \lambda
,n\right) $, $\left( \nu ,n\right) $. All of them have a linear equation of
state. It should be mentioned that all integrable uncharged solutions with $%
\gamma $-law found in Ref. \cite{fortysix} have their charged
generalizations with $p_0\neq 0$ and, hence, possess a boundary. The case $%
n=0$, $p_0=0$ represents CD and arises as an electrification of the trivial
uncharged dust solution (flat spacetime) which is seen from Eqs. (63)-(64).
Its boundary may be put anywhere, since $p=0$ everywhere. The case $n=-1$
generalizes the de Sitter solution into a bunch of new solutions,
characterized by the mass function, subjected to several mild restrictions.
It was completely solved in Sec. IV. The de Sitter solution appears also as
a special case in Sec. X . The option $n=-1/3$ in the uncharged case is a
mark of ESU. It appears in Sec. VII when $b=0$ in Eq. (103). ESU appears
also as the special class $D=0$ in Sec. X. The case $n=-1/5$ is special in
the charged theory too, but its metric is singular as was mentioned in Sec.
VII. The KT solution is generalized in Eq. (155). In the uncharged case
these exhaust the integrable cases of a complicated Abel differential
equation of the second kind. In the charged case it is replaced by a linear
first-order Eq. (18), which has many other solutions, discussed in Sec. VII,
or by a linear second-order Eq. (25), discussed in Sec. X.

From a mathematical viewpoint, the charged case delivers a surprising
variety of equations like those of Euler, Bernoulli, Riccati, Emden-Fowler,
Abel and the hypergeometric equation. The cases containing $\lambda $
quickly lead to special functions whose spectrum is also rich: $%
\mathop{\rm Ei}
$, $cn$, Bessel, hypergeometric, incomplete $\beta $-functions, Jacobi,
Gegenbauer and Chebyshev polynomials, etc. Therefore, a more thorough study
would require computer simulations of special functions, which are easier
than purely numeric simulations.

From a physical viewpoint, the most interesting results are, in our opinion,
the following:

1) The case of constant $T_0^0$ is solved by a simple algorithm involving
algebraic operations, one differentiation and one simple integration upon a
generating function $Y$, which satisfies few simple inequalities, Eqs.
(36)-(38).

2) The charged de Sitter case is completely soluble in terms of the mass or
the charge function, Eqs. (41)-(44). There is a generalization with positive
pressure given by Eqs. (52)-(54).

3) The general solution for a given pressure consists of three contributions
to $z$; from regular CD, from general CD and from the pressure, see Eq.
(56). There is a general CD solution which creates a halo around charged
fluid balls and may postpone their junction to a RN solution.

4) The solutions of Korkina-Durgapal possess realistic charge
generalizations given by Eqs. (88)-(90).

5) Fluids with linear equation of state are integrable for any $n$ as seen
from Eqs. (18), (25), (28), (97)-(100). Elementary solutions may be found by
several simplification techniques. Physically realistic is an infinite
series of models given by Eqs. (104), (105)-(106). The status of the model
given by Eq. (111) is still unknown.

6) The case $\left( \nu ,\rho \right) $ is closely connected to $\left( \nu
,n\right) $ and is much more difficult than its companion with constant $%
T_0^0$, even when the density is constant or zero.

7) The $\left( \rho ,q\right) $ case discussed in Ref. \cite{twentysix} is
incomplete and has errors. It is treated in detail in Sec. IX and its
connection with Ref. \cite{six} is elucidated. It has several realistic
subcases, but electromagnetic mass models all seem to have negative
pressure. A formal, but intriguing parallel is drawn to star models with
simple spatial geometry \cite{fortyfive}.

8) Solutions of the case $\left( \lambda ,n\right) $ with constant $T_0^0$
are expressed in hypergeometric functions, see Eqs. (156), (157). All
degenerate elementary solutions have either negative pressure or
singularities.

A more detailed version of the present paper can be found in Ref. \cite
{Ivanov}.

\end{document}